\documentclass[a4paper]{IEEEtran}

\usepackage{acro}
\usepackage[utf8]{inputenc}
\usepackage[T1]{fontenc}
\usepackage{textcomp}
\usepackage{mathptmx}
\usepackage{fancyvrb}
\usepackage{algorithm}
\usepackage{algpseudocode}
\usepackage{mathtools}
\usepackage{bbm}
\usepackage[flushleft]{threeparttable}	
\usepackage{etoolbox}\AtBeginEnvironment{algorithmic}{\scriptsize} 
\usepackage[usenames,dvipsnames]{xcolor,colortbl}
\usepackage[colorlinks,allcolors=blue]{hyperref}
\usepackage{adjustbox}
\usepackage{amssymb}
\usepackage{pgfplots}
\usepackage{pgfplotstable}
\usepackage{filecontents}
\usepackage{dblfloatfix}    

\usepgfplotslibrary{colorbrewer}
\usetikzlibrary{pgfplots.statistics, pgfplots.colorbrewer} 

\pgfkeys{/pgf/number format/.cd,1000 sep={\,},fixed,precision=3}

\makeatletter
\pgfplotsset{
	boxplot/hide outliers/.code={
		\def\pgfplotsplothandlerboxplot@outlier{}%
	}
}
\makeatother

\newcolumntype{a}{>{\columncolor{Cyan!10!white}}c}
\newcolumntype{b}{>{\columncolor{Gray!10!white}}c}
\newcolumntype{w}{>{\columncolor{red!10!white}}c}
\newcolumntype{x}{>{\columncolor{green!10!white}}c}
\newcolumntype{y}{>{\columncolor{blue!10!white}}c}
\newcolumntype{z}{>{\columncolor{yellow!10!white}}c}

\usepgfplotslibrary{units}
\usetikzlibrary{arrows, calc, spy, patterns, backgrounds,decorations.pathreplacing}

\pgfplotsset{
	my ybar legend/.style={
		legend image code/.code={
			\draw [##1] (0cm,-0.6ex) rectangle +(1.75em,1.1ex);
		},
	},
}

\makeatletter
\pgfdeclarepatternformonly[\hatchdistance,\hatchthickness]{flexible hatch}
{\pgfqpoint{0pt}{0pt}}
{\pgfqpoint{\hatchdistance}{\hatchdistance}}
{\pgfpoint{\hatchdistance-1pt}{\hatchdistance-1pt}}%
{
	\pgfsetcolor{\tikz@pattern@color}
	\pgfsetlinewidth{\hatchthickness}
	\pgfpathmoveto{\pgfqpoint{0pt}{0pt}}
	\pgfpathlineto{\pgfqpoint{\hatchdistance}{\hatchdistance}}
	\pgfusepath{stroke}
}

\makeatletter
\pgfplotsset{
	discontinuous line/.code={
		\pgfkeysalso{mesh, shorten <=#1, shorten >=#1,
			legend image code/.code={
				\draw [##1, shorten <=0cm] (0cm,0cm) -- (0.3cm,0cm);
				\draw [only marks] plot coordinates {(0.3cm,0cm)};
				\draw [##1, shorten >=0cm] (0.3cm,0cm) -- (0.6cm,0cm);
		}}
		\def\pgfplotsplothandlermesh@VISUALIZE@std@fill@andor@stroke{%
			\pgfplotspatchclass{\pgfplotsplothandlermesh@patchclass}{fill path}%
			\pgfplotsplothandlermesh@definecolor
			\pgfusepath{stroke}
			\pgfplotsplothandlermesh@show@normals@if@configured
		}%
	},
	discontinuous line/.default=1.5mm
}
\makeatother

\usepackage[firstinits=true,
uniquename=false,
uniquelist=false,
hyperref=auto,
maxbibnames=4,
maxcitenames=2,
style=numeric,
citestyle=numeric,
backref=false,
natbib=true,
style=trad-abbrv,
doi=true,
backend=biber]{biblatex}
\usepackage{booktabs}
\usepackage{sidecap}
\usepackage{graphicx, caption, subcaption}
\usepackage{relsize}
\usepackage[shrink=20]{microtype}
\usepackage{csquotes}
\usepackage{listings}
\usepackage{algpseudocode}
\usepackage[curves]{struktex}
\usepackage{mathrsfs}
\usepackage{multicol}

\usepackage{makeidx}  
\usepackage{icomma}
\usepackage{color}
\usepackage{array}
\usepackage{multirow}
\usepackage{layouts}
\usepackage{mathptmx}
\usepackage{breakurl}
\usepackage{comment}
\usepackage{ifthen}

\usepackage{amsmath}
\usepackage{amsfonts}
\usepackage{relsize}
\usepackage{xspace}

\usepackage{todonotes}
\usepackage{setspace}
\usepackage{cleveref}
\usepackage[binary-units=true]{siunitx}

\newcommand{\acrodef}[2]{\DeclareAcronym{#1}{short={#1},long={#2}}}
\acrodef{AABB}{axis-aligned bounding box}
\acrodef{API}{application programming interface}
\acrodef{ASIC}{application specific integrated circuit}
\acrodef{AST}{abstract syntax tree}
\acrodef{AVX}{advanced vector extensions}
\acrodef{BRAM}{block RAM}
\acrodef{CB}{compute-bound}
\acrodef{CER}{communication-to-execution ratio}
\acrodef{CG}{conjugate gradient}
\acrodef{ChebFD}{Chebyshev filter diagonalization}
\acrodef{CL}{cache line}
\acrodef{CoD}{cluster-on-die}
\acrodef{CPI}{cycles per instruction}
\acrodef{CPU}{central processing unit}
\acrodef{CUDA}{compute unified device architecture}
\acrodef{CST}{concrete syntax tree}
\acrodef{DP}{double precision}
\acrodef{DPM}{delay propagation mechanism}
\acrodef{DOF}{degree of freedoms}
\acrodef{DOM}{delay overlapping mechanism}
\acrodef{DPOM}{delay propagation and overlapping mechanisms}
\acrodef{DSL}{domain-specific language}
\acrodef{DVFS}{dynmic voltage frequency scaling}
\acrodef{ECM}{execution-cache-memory}
\acrodef{FD}{finite difference}
\acrodef{FEM}{finite element method}
\acrodef{FFC}{FEniCS Form Compiler}
\acrodef{FFT}{Fast Fourier transform}
\acrodef{FIFO}{first in first out}
\acrodef{FLOPS}{floating point operations per second}
\acrodef{FMA}{fused multiply-add}
\acrodef{FP}{floating-point}
\acrodef{FPGA}{field-programmable gate array}
\acrodef{FV}{finite volume}
\acrodef{GMRES}{generalized minimal residual}
\acrodef{GPU}{graphics processor unit}
\acrodef{GS}{Gauss-Seidel}
\acrodef{GUI}{graphical user interface}
\acrodef{HPCG}{High Performance Conjugate Gradient}
\acrodef{HDL}{hardware description language}
\acrodef{HHG}{hierarchical hybrid grid}
\acrodef{HLS}{high-level synthesis}
\acrodef{HPC}{high-performance computing}
\acrodef{IACA}{Intel Architecture Code Analyzer}
\acrodef{IP}{intellectual property}
\acrodef{ISA}{instruction set architecture}
\acrodef{ITAC}{Intel trace analyzer and collector}
\acrodef{IR}{intermediate representation}
\acrodef{JIT}{just-in-time}
\acrodef{KPM}{Kernel Polynomial Method}
\acrodef{LC}{Layer Condition}
\acrodef{LFA}{local Fourier analysis}
\acrodef{LBM}{Lattice Boltzmann}
\acrodef{LLC}{last-level cache}
\acrodef{LoC}{lines of code}
\acrodef{LZR}{Leibniz Supercomputing Centre}
\acrodef{MB}{memory-bound}
\acrodef{MC}{memory controller}
\acrodef{MPI}{Message Passing Interface}
\acrodef{NDG}{nodal discontinuous Galerkin}
\acrodef{NDGTD}{nodal discontinuous Galerkin time domain}
\acrodef{NIC}{network interface controller}
\acrodef{OMP}{OpenMP}
\acrodef{NT}{non-temporal}
\acrodef{NUMA}{non-uniform memory access}
\acrodef{OS}{operating system}
\acrodef{OSACA}{Open-Source Architecture Code Analyzer}
\acrodef{P2P}{peer-to-peer}
\acrodef{PDE}{partial differential equation}
\acrodef{RAPL}{running average power limit}
\acrodef{PGAS}{partitioned global address space}
\acrodef{PPnR}{post place and route}
\acrodef{PPS}{processes per socket}
\acrodef{QDR}{quad data rate}
\acrodef{RAM}{random access memory}\acuse{RAM}
\acrodef{RBGS}{red-black Gauss-Seidel}
\acrodef{RDMA}{remote direct memory access}
\acrodef{RHS}{right-hand side}
\acrodef{RRZE}{Regional Computer Center Erlangen} 
\acrodef{RTL}{register transfer level}
\acrodef{SHM}{shared memory}
\acrodef{SPIR}{standard portable intermediate representation}
\acrodef{SPL}{software product lines}
\acrodef{SIMD}{single instruction, multiple data}
\acrodef{SMP}{symmetric multiprocessing}
\acrodef{SMT}{simultaneous multithreading}
\acrodef{SP}{single precision}
\acrodef{SSE}{streaming SIMD extensions}
\acrodef{STL}{Standard Template Library}
\acrodef{TDP}{thermal design power}
\acrodef{TLB}{translation lookaside buffer}
\acrodef{TPDL}{target platform description language}
\acrodef{UFS}{Uncore frequency scaling}
\acrodef{WF}{wavefront}
\acrodef{XML}{eXtensible Markup Language}

\newcommand{\CPP}{C\nolinebreak[4]\hspace{-.05em}\raisebox{.23ex}{\relsize{-1}{++}}}

\newif\iftitle
\titletrue

\newcommand{\bq}{\begin{equation}}
\newcommand{\eq}{\end{equation}}
\newcommand{\bytes}{\mbox{B}}
\newcommand{\byte}{\mbox{byte}}
\newcommand{\second}{\mbox{s}}
\newcommand{\MS}{\mbox{ms}}

\newcommand{\flop}{\mbox{flop}}
\newcommand{\flops}{\mbox{flops}}

\newcommand{\cycles}{\mbox{cy}}

\newcommand{\FS}{\mbox{\flop/\second}}

\newcommand{\GBS}{\mbox{GB/\second}}

\newcommand{\GB}{\mbox{GB}}

\newcommand{\eos}{~.}
\newcommand{\cma}{~,}

\newcommand{\Rlm}{Roof{}line model\xspace}
\newcommand{\Rl}{Roof{}line\xspace}

\newcommand{\mrn}[1]{\mathrm{\RN{#1}}}

\SaveVerb{Emmy}.ClusterA.
\SaveVerb{Meggie}.ClusterB.
\SaveVerb{HazelHen}.Hazel Hen.
\SaveVerb{SuperMUCNG}.SuperMUC-NG.
\SaveVerb{Hawk}.Hawk.
\SaveVerb{MPIcommsize}.MPI_comm_size.
\SaveVerb{MPIwaitall}.MPI_Waitall.
\SaveVerb{MPIwait}.MPI_Wait.
\SaveVerb{MPIisend}.MPI_Isend.
\SaveVerb{MPIsend}.MPI_Send.
\SaveVerb{MPIirecv}.MPI_Irecv.
\SaveVerb{MPIrecv}.MPI_Recv.
\SaveVerb{MPIsendrecv}.MPI_Sendrecv.
\SaveVerb{MPIbarrier}.MPI_Barrier.
\SaveVerb{MPIallreduce}.MPI_Allreduce.
\SaveVerb{MPIreduce}.MPI_Reduce.
\SaveVerb{MPIalltoall}.MPI_Alltoall.
\SaveVerb{MPIallgather}.MPI_Allgather.
\SaveVerb{MPIgather}.MPI_Gather.
\SaveVerb{MPIscatter}.MPI_Scatter.
\SaveVerb{MPIbcast}.MPI_Bcast.
\SaveVerb{MPI}.MPI.
\SaveVerb{OpenMP}.OpenMP.
\SaveVerb{MPIOpenMP}.MPI+OpenMP.
\SaveVerb{NUMA}.NUMA.
\SaveVerb{STREAM}.STREAM.
\SaveVerb{Cosine}.Cosine.
\SaveVerb{DIV}.DIV.
\SaveVerb{vdivpd}.vdivpd.
\SaveVerb{SpMV}.SpMV.
\SaveVerb{static}.static.
\SaveVerb{LIKWID}.ToolA.
\SaveVerb{likwidperfctr}.featureA.
\SaveVerb{Chrono}.Chrono.
\SaveVerb{ccNUMA}.ccNUMA.
\SaveVerb{BLAS1}.BLAS-1.
\SaveVerb{BLAS2}.BLAS-2.

\definecolor{myblue}{RGB}{37,165,203}
\definecolor{myred}{RGB}{175,32,67}

\colorlet{ghcolor}{ProcessBlue}
\colorlet{aycolor}{RubineRed}

\makeatletter
\renewcommand{\todo}[2][]{\@todo[caption={#2},#1]{\begin{spacing}{0.5}\fontfamily{phv}\fontseries{mc}\selectfont{#2\vspace{-1em}}\end{spacing}}}
\makeatother

\acuse{CPU}

\newif\ifblind
\blindfalse
  
\begin{document}
\title{An analytic performance model for overlapping execution of memory-bound loop kernels on multicore CPUs}
 
\ifblind
\author{Authors omitted for double-blind review process}
\else

\author{
	\IEEEauthorblockN{Ayesha Afzal, Georg Hager, Gerhard Wellein}\\
	\IEEEauthorblockA{\ Department of High Performance Computing\\ Erlangen Regional Computing Center\\ Friedrich-Alexander-Universit\"at Erlangen-N\"urnberg, Germany\\ \{ayesha.afzal,georg.hager,gerhard.wellein\}@fau.de}
}
\fi

\maketitle
\pgfkeys{/pgf/number format/.cd,1000 sep={\,}}

\begin{abstract}
  Complex applications running on multicore processors show a rich 
  performance phenomenology. The growing number of cores per ccNUMA
  domain complicates performance analysis of memory-bound code since
  system noise, load imbalance, or task-based programming models can
  lead to thread desynchronization. Hence, the simplifying assumption that
  all cores execute the same loop can not be upheld. Motivated by
  observations on plain and modified versions of the HPCG benchmark,
  we construct a performance model of execution of
  memory-bound loop kernels. It can predict the memory bandwidth
  share per kernel on a memory contention domain depending on the
  number of active cores and which other workload the kernel is
  paired with.  The only code features required are
  the single-thread cache line access frequency per kernel, which is
  directly related to the single-thread memory bandwidth, and its
  saturated bandwidth.  It can either be
  measured directly or predicted using the Execution-Cache-Memory (ECM)
  performance model. The computational intensity of the kernels and
  the detailed structure of the code is of no
  significance.  We validate our model on Intel Broadwell, Intel Cascade
  Lake, and AMD Rome processors pairing various streaming and stencil
  kernels. The error in predicting the bandwidth share per kernel
  is less than 8\%.
  
\end{abstract}

\section{Introduction}

With the number of cores and the peak performance of modern multicore
chips still growing, the memory bandwidth bottleneck is becoming more
severe. Many algorithms in computational science are based on building
blocks that show memory-bound behavior, i.e, whose performance does
not scale but saturate with respect to the number of active cores when
running on a memory contention domain (usually a ccNUMA domain).
Performance saturation due to bandwidth saturation should be regarded
as a sign that a code is fast enough to address an architectural
bottleneck, so it is not a bad thing in general. It also opens
a clear optimization path via the reduction of data transfers.

Beyond the simple saturation pattern, however, memory-bound code
exhibits other, more interesting phenomenology. As was shown in~\cite{AfzalHW20},
bulk-synchronous barrier-free MPI programs can show \emph{desynchronization}
on a contention domain, i.e., processes move away from the initial
lockstep state into a state where execution of loops kernels overlaps
with communication or idleness, leading to automatic communication hiding.
This can be provoked by a deliberate injection of delays, but it
can also occur automatically by natural system noise and
small load imbalances. 

Although it is often observed in programs that have a significant
communication overhead, desynchronization is not limited
to this scenario. In fact, barrier-free code (and also
modern, task-based programming models) allows for concurrent
execution of code with very different characteristics
on a contention domain. In this paper we investigate
a specific scenario: concurrent execution of two
different loop kernels on $n$ cores each. We construct
a performance model that can describe accurately the
memory bandwidth share that each kernel attains. The model
is thus able to predict whether desynchronization of back-to-back
parallel kernels and the ensuing overlap will speed up or slow
down the execution of threads compared to a purely
homogeneous execution. It can also predict if
the mutual overlap will amplify or reduce the
desynchronization effect.

This paper is organized as follows. The rest of this section covers
related work and a motivational example (HPCG). In Sect.~\ref{sec:exp}
we describe the experimental setup and methodology, and
Sect.~\ref{sec:metrics} defines code and performance metrics. The
performance model is developed in~\ref{sec:model} and validated in
Sect.~\ref{sec:validation}, where we also discuss the connection to
desynchronization and close the loop to the initial HPCG example.
We give a summary and an outlook to
future work in Sect.~\ref{sec:conclusion}.

\subsection{Motivation and related work}\label{sec:motivation}
\renewcommand{\thefigure}{2}
\begin{figure*}[!b]
	\begin{tikzpicture}
		\put(-514.5,0){\includegraphics[width=.98\textwidth,height=0.051\textheight]{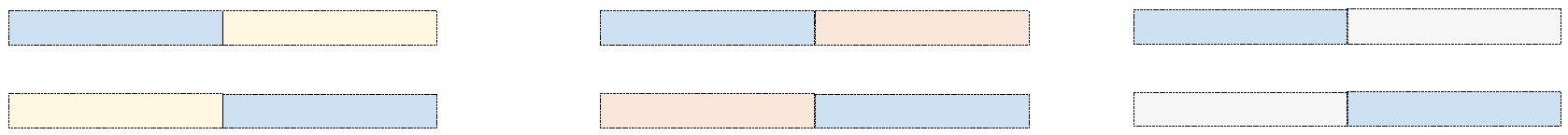}
		\node [font=\small] at (-16.8,0.97){Exec\_A0}; 
		\node [font=\small] at (-16.8,0.22){Exec\_B1}; 
		\node [font=\small] at (-14.3,0.97){Exec\_B0}; 
		\node [font=\small] at (-14.3,0.22){Exec\_A1}; 
		\node [font=\small] at (-9.9,0.97){Exec\_A0}; 
		\node [font=\small] at (-9.9,0.22){Comm\_B1}; 
		\node [font=\small] at (-7.4,0.97){Comm\_B0}; 
		\node [font=\small] at (-7.4,0.22){Exec\_A1}; 
		\node [font=\small] at (-3.8,0.97){Exec\_A0}; 
		\node [font=\small] at (-3.8,0.22){Idle\_B1}; 
		\node [font=\small] at (-1.3,0.97){Idle\_B0}; 
		\node [font=\small] at (-1.3,0.22){Exec\_A1}; 
		\node [font=\small] at (-15.55,-1.2){a1. Identical kernels: Exec\_B1 := Exec\_A1};
		\node [font=\small] at (-15.55,-1.6){a2. Different kernels: Exec\_A $\neq$ Exec\_B}; 
		\node [font=\small] at (-8.65,-1.2){b1. Latency-limited communication};
		\node [font=\small] at (-8.65,-1.6){b2. Bandwidth-limited communication};
		\node [font=\small] at (-18,0.97){P0}; 
		\node [font=\small] at (-18,0.22){P1};
		\draw [thick, ->] (-18,-0.3) -- (-13,-0.3); 
		\draw [thick, ->] (-11.2,-0.3) -- (-6.2,-0.3);
		\draw [thick, ->] (-5,-0.3) -- (0,-0.3);
		\node [font=\small] at (-15.55,-0.65){Time}; 
		\node [font=\small] at (-8.65,-0.65){Time}; 
		\node [font=\small] at (-2.55,-0.65){Time}; 
		\node [font=\small] at (-2.55,-1.6){(c)}; 
	}
	\end{tikzpicture}
	\caption{Characterization of concurrency scenarios for
		bandwidth-limited computational kernels in parallel
		programs. ``P0'' and ``P1'' denote multi-threaded processes
		or groups of MPI processes within a contention domain.  Each
		thread or process in a group executes the same code on
		different data.  (a1, a2) Overlapping memory-bound loop kernels
		of different (or same) characteristics, (b1, b2) overlapping loop kernel
		execution with latency- or bandwidth-bound communication, and (c) overlapping loop
		kernel execution with idleness or code that addresses
		nonshared or scalable resources.
		\label{fig:Overlap}}
\end{figure*}
\renewcommand{\thefigure}{1}
\begin{figure}[bt]
  \input{figures/fig3_HPCG.tex}
  \caption{MPI-only HPCG traces on one ccNUMA domain of Intel
    Broadwell (BDW-2) and Cascade Lake (CLX) at a problem size of
    $160^3$ per process. Colors denote SymGS (orange), DDOT2 (pink),
    and \texttt{MPI\_Allreduce} (green). Top panels in (a) and (b):
    timeline snippets from around a time stamp of 1628\,s ($x$
    axes show offsets from this point). Bottom panels in (a) and (b):
    same data but with ranks sorted by the starting time of
    the DDOT2 kernel.
    (c) DDOT2 runtime per rank, same sorting as in bottom
    panels of (a) and (b).    
  \label{fig:HPCG}}
\end{figure}

This investigation is based on prior work by Afzal et
al.~\cite{AfzalEuroMPI19Poster,AfzalHW19,AfzalHW20}, who studied idle
wave propagation and computational wave formation in parallel
programs on a phenomenological level. An analytic model of
overlapping computational kernels and communication time on a contention domain
was lacking, however. Alappat et al.~\cite{Alappat:ISC20}
observed the consequences of desynchronization in  the
context of the well-known HPCG
benchmark:\footnote{\url{{https://www.hpcg-benchmark.org/}}}
While validating their \Rl\ model for the MPI-parallel
HPCG, they observed that the
DDOT2 (dot product \texttt{s+=a[i]*b[i]}) kernels were in fact faster than what the
local memory bandwidth would allow for. The authors attributed
this to process desynchronization during the preceding
sparse matrix-vector multiplication (SpMV) kernel and assumed
that MPI processes that started the DDOT2 kernel early could
benefit from immediate cache reuse, which was backed by
a measured computational intensity that was higher than
expected. However, it was unclear whether cache reuse was the
only reason for the observed elevated DDOT2 performance. 

The HPCG algorithm comprises the following kernels: six of BLAS-1 type
(two DDOT2, one DDOT1 [\texttt{s+=a[i]*a[i]}] and three DAXPY [\texttt{a[i]+=s*b[i]}])
and one SpMV.  Additionally, the multigrid preconditioner compromises five
kernels: restriction, prolongation, SpMV, and 
two symmetric Gauss-Seidel routines (SymGS) as pre-
and post-smoothers, each with a forward and a backward sweep, for
coarsening and refinement. HPCG has two types of MPI
communication: a global collective operation (\verb.MPI_Allreduce.)
after each vector dot product
and nonblocking point-to-point communication in SpMV and SymGS.

Figure~\ref{fig:HPCG}(a) (top) shows a snippet from
a timeline diagram of the HPCG benchmark (problem size $160^3$
per process) on the nine cores of one ccNUMA domain of an Intel
Broadwell processor (see Sect.~\ref{sec:exp} for hardware details
and software setup).
The colors encode the SymGS kernel (orange), the DDOT2 kernel (pink),
and the \verb.MPI_Allreduce. call (green). Note that this is
only a part of a single iteration; all effects described here
occur in every iteration of the algorithm.

The runtime of
the SymGS kernel is about 20 times longer than that of DDOT2 here,
so the clearly visible desynchronization effect has a minor influence
on the SymGS performance. However, the execution of DDOT2
is strongly out of sync across processes. ``Early'' ranks still
overlap with SymGS running on other cores, while ``late'' ranks overlap
with waiting time (i.e., idleness) in \verb.MPI_Allreduce..
The former must compete for memory bandwidth with SymGS and thus take
longer to execute, while
the latter can utilize more bandwidth per core, making them execute
faster. This can be seen better in the lower part of Fig.~\ref{fig:HPCG}(a),
where we have sorted the MPI ranks according to the starting time
of the DDOT2 kernel from early (bottom) to late (top). The
late starters clearly take less time than the early starters, although
the latter are those with cache reuse potential.
In Fig.~\ref{fig:HPCG}(b), the experiment was repeated on a 20-core
Intel Cascade Lake processor. The effect 
is less pronounced here since the DDOT2 runtime is now significantly longer than the
desynchronization time scale. There is hence less opportunity
for DDOT2 to overlap with  \verb.MPI_Allreduce..

Figure~\ref{fig:HPCG}(c) shows the runtime of the DDOT2 kernel per rank
in both experiments, using the same sorting as in (a) and (b).
The runtimes are monotonically decreasing in both cases, showing
clearly that late starters have better performance.
In summary, the desynchronization of the SymGS kernel, in itself a
negligible effect, leads to some cores executing the DDOT2 kernel
faster due to overlapping with idleness in the
\verb.MPI_Allreduce. function. This is the real reason for the
``faster-than-light'' performance observed in~\cite{Alappat:ISC20}.

In the broader context of desynchronized execution of barrier-free
bulk-synchronous code, the general question arises how much bandwidth
and thus performance can be achieved per core if different
computational kernels are executed concurrently by cores on the same
contention domain, if part of the cores are idle, or if they spend
time communicating. Figure~\ref{fig:Overlap} shows the range of
possible scenarios.

While scenario (b), i.e., overlapping of loop kernel execution with
communication, was covered in~\cite{AfzalHW20}, and scenario (c)
is well described by the Execution-Cache-Memory performance model
for multicore processors~\cite{Hofmann:2020:SF}, scenario (a)
was not studied before. 

\renewcommand{\thefigure}{3}
\begin{figure}[bt]
	\input{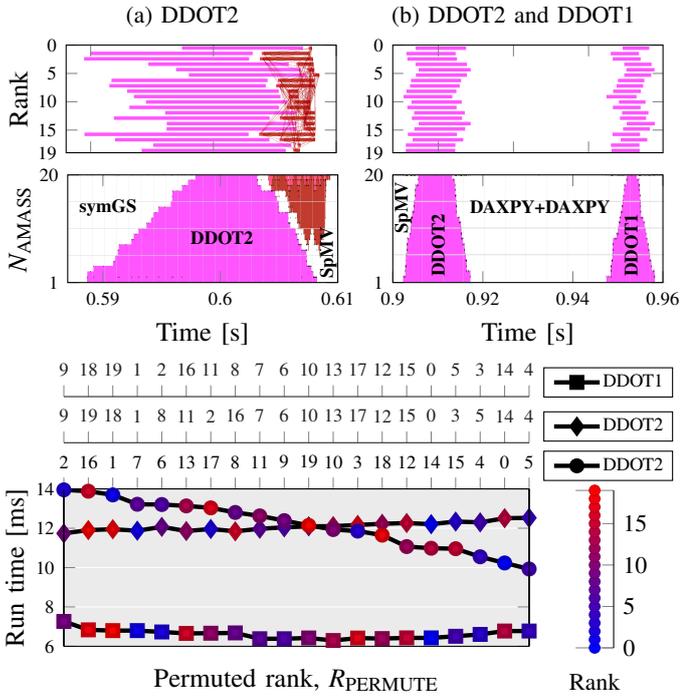}
	\caption{Top panels of (a) and (b): Timeline snippets of a modified
		HPCG variant (missing any reduction operations) on an Intel CLX
		socket at a problem size of $160^3$ per process,
		showing (a) DDOT2 between SymGS and SpMV
		and (b) DDOT2 and DDOT1 with two DAXPYs in between.
		The bottom panels of (a) and (b) show quantitative timelines
		of the number of ranks concurrently executing the DDOT kernels.
		Dark red color in (a) denotes time spent in
		\texttt{MPI\_Wait} for nonblocking point-to-point communication during SpMVM.
		(c) shows time spent in the DDOTx kernels per core vs.\ permuted
		rank (permutation as in Fig.~\ref{fig:HPCG}).
		\label{fig:Modified_HPCG}}
\end{figure}

To further set the stage and motivate the relevance of the bandwidth
sharing phenomenon, we conducted further experiments with a modified
variant of HPCG that is identical to the original except for
the lack of \verb.MPI_Allreduce. calls. This allows for
desynchronized states to survive for a long time, since
no other global MPI operations are used in HPCG. We also
executed a single iteration of the benchmark only.
Figures~\ref{fig:Modified_HPCG}
(a) and (b) shows timeline snippets of the modified HPCG on the same
Intel Cascade Lake CPU as before. In Fig.~\ref{fig:Modified_HPCG}(a)
the DDOT2 kernel is sandwiched between a SymGS and an SpMV, while
in Fig.~\ref{fig:Modified_HPCG}(b) we show a broader view of the
two other DDOT kernels, with two DAXPY operations in between.
As all global reductions were removed, the DDOTs are allowed
to overlap with subsequent kernels.

In the bottom panels of (a) and (b) we show quantitative timelines
that tell how many ranks execute the DDOT kernel at each point in time.
This makes it easy to spot phases of homogeneous (i.e, single-kernel)
execution and desynchronization.
Qualitatively different
behavior across the different DDOT kernels can be observed:
In (a), where the tail end of the DDOT2 execution overlaps with
a subsequent SpMV and its significant \verb.MPI_Wait. time,
the spread in
endpoints of DDOT2 execution is smaller than the spread
in their starting points. This is plausible because, as described
earlier, early starters of DDOT2 take longer while late starters
are faster. The effect can be quantified by a
negative skewness parameter of the accumulated DDOT2 time
distribution of $-0.27\,\MS$. The second DDOT2, however, is sandwiched
between SpMV and DAXPY, so any overlap with idleness is ruled
out. Instead, the interaction with the low-intensity DAXPY
code seems to boost the desynchronization effect, i.e., the tail
end of the accumulated DDOT2 time distribution is longer than
the front end (skewness $0.42\,\MS$). The last DDOT, which is
actually a vector norm computation (DDOT1), shows a similar
characteristic with an even larger positive skewness of $1.0\,\MS$.

The skewness parameterizes the asymmetry in the distribution of a
random variable, but the timeline data from the HPCG is not normally
distributed even if the skewness is zero. However, we can use the
skewness here to identify an important property of barrier-free
parallel loop kernel execution: Negative skewness indicates
\emph{resynchronization}, while positive skewness indicates
\emph{desynchronization}. Which of the two occurs is obviously a
function of which other workloads the kernel is embedded in. Note that
the described effects in HPCG, even in our synchronization-free
variant, are small in absolute numbers since the HPCG runtime is
dominated by the multigrid and SpMV kernels. However, our observations
open interesting questions towards the onset and mitigation of
desynchronization effects in barrier-free memory-bound MPI programs. The goal of this
paper is to establish a performance model for loop kernels of
different characteristics running concurrently on a contention domain
that can describe this scenario with sufficient accuracy.
As an aditional benefit, such a model can also predict the
dynamics of task-based programming models on memory-bound code.

\subsection{Contributions}
This paper makes the following new contributions.
\renewcommand{\thefigure}{4}
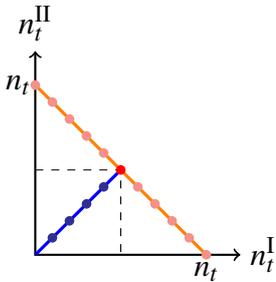
\begin{SCfigure}[][!tb]
	\begin{tikzpicture}[scale=0.9]
    \draw [<->,thick,font=\large] (0,3) node (yaxis) [above] {$n_t^{\mathrm{\RN{2}}}$}
        |- (3,0) node (xaxis) [right] {$n_t^{\mathrm{\RN{1}}}$};

    \draw [
    very thick,
    blue
    ] (0,0) coordinate (a_1) -- (1.25,1.25) coordinate (a_2);
    \draw [
    very thick,
    orange
    ] (0,2.5) coordinate (b_1) -- (2.5,0) coordinate (b_2);

    \coordinate (c) at (intersection of a_1--a_2 and b_1--b_2);
    
    \coordinate (d) at (0.25,0.25);
    
    \coordinate (e) at (0.5, 0.5);
    
    \coordinate (f) at (0.75,0.75);
    
    \coordinate (g) at (1, 1);
    
    \coordinate (h) at (0.25,2.25);

	\coordinate (i) at (0.5, 2);

	\coordinate (j) at (0.75,1.75);

	\coordinate (k) at (1, 1.5);

	\coordinate (m) at (1.5, 1);

	\coordinate (n) at (1.75,0.75);

	\coordinate (o) at (2, 0.5);
	
	\coordinate (p) at (2.25, 0.25);
	
	\coordinate (q) at (2.5,0);
	
	\coordinate (r) at (0,2.5);
	
    \draw[dashed] (yaxis |- c) node[left] {}
        -| (xaxis -| c) node[below] {};
    
    \node [font=\large] at (2.5,-0.25) {$n_t$};
    \node [font=\large] at (-0.25,2.5) {$n_t$};

    \fill[red] (c) circle (2pt);
   	\fill[Blue] (d) circle (2pt);
    \fill[Blue] (e) circle (2pt);
    \fill[Blue] (f) circle (2pt);
    \fill[Blue] (g) circle (2pt);
    \fill[Salmon] (h) circle (2pt);
	\fill[Salmon] (i) circle (2pt);
	\fill[Salmon] (j) circle (2pt);
	\fill[Salmon] (k) circle (2pt); 
	\fill[Salmon] (m) circle (2pt);
	\fill[Salmon] (n) circle (2pt);
	\fill[Salmon] (o) circle (2pt); 
	\fill[Salmon] (p) circle (2pt);
	\fill[Salmon] (q) circle (2pt);
	\fill[Salmon] (r) circle (2pt);
\end{tikzpicture} 
	\caption{Parameter space covered
		w.r.t.\ number of cores sharing a bottleneck (dots).
		$n_t^I$, $n_t^{II}$: number of cores running kernel $I$ and $II$;
		$n_t$: number of cores on ccNUMA domain.
                Orange: domain fully occupied; blue: symmetrical scaling.
		%
		\label{fig:threads}}
\end{SCfigure}
\begin{table*}[!tb]
	\centering
	\begin{adjustbox}{width=0.98\textwidth}
		\begin{threeparttable}
			\caption{Key hardware and software specifications of systems.} 
			\setlength\extrarowheight{-0.7pt}
\setlength\tabcolsep{2pt}
\arrayrulecolor{Blue}
\begin{tabular}[fragile]{c>{~~}lwxyz}
	\toprule
	\rowcolor[gray]{0.9}
	&Systems  & BDW-1   & BDW-2 & CLX & Rome\\
	\midrule
	& Processor  & Broadwell EP  & Broadwell EP  & Cascade Lake SP & Zen  \\     
	& Model      & Intel Xeon E5-2630 v4   &  Intel Xeon E5-2697 v4           & Intel Xeon Gold 6248 & ARM Epyc 7451\\
	
	&Base clock speed &\SI{2.2}{\giga \Hz}        & \SI{2.3}{\giga \Hz} & \SI{2.5}{\giga \Hz}& \SI{2.35}{\giga \Hz}\\
	
	\cellcolor{Cyan!10!white}&Cores per NUMA domain (SMT)    & 10 (20)    &    18 (36)        & 20 (40)& 8 (16)\\
	
	\cellcolor{Cyan!10!white}&Private L1 size & 32 ($\times$10) \si{\kibi \byte}    &     32 ($\times$18) \si{\kibi \byte}        & 32 ($\times$20) \si{\kibi \byte}& 32 ($\times$8) \si{\kibi \byte} \\
	
	\cellcolor{Cyan!10!white}&Private L2 size & 256 ($\times$10) \si{\kibi \byte}    &     256 ($\times$18) \si{\kibi \byte}        & 1048 ($\times$20) \si{\kibi \byte}& 512 ($\times$8) \si{\kibi \byte} \\
	
	\cellcolor{Cyan!10!white}&Shared LLC size & \SI{25}{\mebi \byte} (10$\times$\SI{2.5}{\mebi \byte})    &    \SI{45}{\mebi \byte} (18$\times$\SI{2.5}{\mebi \byte})          &\SI{27.5}{\mebi \byte} (20$\times$\SI{1.375}{\mebi \byte})& \SI{8}{\mebi \byte} (per 4 cores)\\
	
	\cellcolor{Cyan!10!white}&Memory per node (type)&  \SI{64}{\gibi \byte} (DDR4)    & \SI{384}{\gibi \byte} (DDR4) & \SI{128}{\gibi \byte} (DDR4)& \SI{128}{\gibi \byte} (DDR4)\\
	
%

	\cellcolor{Cyan!10!white}&LD/ST throughput (SIMD)  &   2/1 (AVX2/FMA3)  & 2/1   (AVX2/FMA3)             & 2/1 (AVX-512/FMA3)& 2/1 (AVX2/FMA3)\\

	\cellcolor{Cyan!10!white}&L1 $\rightleftarrows$  L2 bandwidth 
	      & \SI{64}{B\per\cycles} 
	&  \SI{64}{B\per\cycles}  & \SI{64}{B\per\cycles}& 32+32 \si{B\per\cycles}\\
	
	\cellcolor{Cyan!10!white}&L2 $\rightleftarrows$ LLC bandwidth 
	       & 32 \si{B\per\cycles} 
	&   32 \si{B\per\cycles}             & 16+16 \si{B\per\cycles} & 32 \si{B\per\cycles}\\ 

	\cellcolor{Cyan!10!white}& LLC organization 
	& Inclusive 
	&   Inclusive            & Exclusive & Exclusive\\
		
	& Victim caches   
	& No     
	&  No            &  LLC &  LLC\\
	
	& El. transfers   
	& Non-overlapping     
	&  Non-overlapping            & Non-overlapping & Overlapping\\

	
	\multirow{-14}{*}{\rotatebox{90}{ \textbf{Micro-architecture}}} &Theor. MEM bandwidth & \SI{68.3}{\giga \byte / \second}& \SI{76.8}{\giga \byte / \second} &  \SI{140.8}{\giga \byte / \second}& \SI{170.6}{\giga \byte / \second}\\
	
	
	\midrule
	\cellcolor{Cyan!10!white}&Compiler    & Intel \CPP{} v2019.5.281         & Intel \CPP{} v2019.5.281     &  Intel \CPP{} v2019.5.281  &  Intel \CPP{} v2019.5.281  \\
	\cellcolor{Cyan!10!white}&Optimization flags & -O3 -xHost  & -O3 -xHost & -O3 -qopt-zmm-usage=high & -O3 -xHost \\
	\cellcolor{Cyan!10!white}&(SIMD) & -xAVX &-xAVX&-xCORE-AVX512 & -mavx2\\
	\cellcolor{Cyan!10!white}&Message passing library & Intel \verb.MPI. v2019u5         & Intel \verb.MPI. v2019u5     & Intel \verb.MPI. v2019u5   & Intel \verb.MPI. v2019u5  \\
	\multirow{-4}{*}{\rotatebox{90}{\cellcolor{Cyan!10!white} \textbf{Software}}}&Operating system    & CentOS Linux v7.7.1908      &     Ubuntu 18.04.3       & Ubuntu 18.04.3& Ubuntu 18.04.3\\  
	
	\midrule
	\cellcolor{Cyan!10!white}&\verb.ITAC.    & v2019u4         & v2019u4    & v2019u4& v2019u4\\
	\multirow{-2}{*}{\rotatebox{90}{\cellcolor{Cyan!10!white} \textbf{Tools}}}&\verb.LIKWID.   & v5.0.1         & v5.0.1   & v5.0.1 & v5.0.1\\
	\bottomrule
\end{tabular}

			%
			\label{tab:system}
		\end{threeparttable}
	\end{adjustbox}
\end{table*}
(i) We identify the influence factors that govern the bandwidth
  share for individual threads in a ``hybrid execution'' setting,
  specifically two (groups of) threads running loops with different
  characteristics on a memory bandwidth contention domain:
  the ``memory request fraction''
  to the memory interface and the saturated memory bandwidth of
  the different kernels. 
(ii) We employ ideas from the ECM performance model to
  predict the bandwidth share of different
  kernels on a contention domain.
(iii) We validate the model for a set of 30 kernel pairings on
  four x86 architectures.
  Despite sigificant differences in hardware properties across
  systems, the modeling error is below 8\% globally and
  below 5\% for 75\% of cases.

Figure~\ref{fig:threads} visualizes the parameter space covered in this
work with respect to numbers of threads sharing a contention domain.
We only combine two distinct kernels, either filling the domain
completely (orange dots) or scaling with equal core counts towards
full saturation (blue dots). Other combinations were ignored
for brevity although they can be described by the model just as well.

\begin{table*}[!tb]
	\centering
	\begin{threeparttable}
		\caption{Key specifications of computational kernels used in the experiments.
			All arrays and scalars hold double-precision floating point data.
			In all cases with write access, standard stores and write-back caches are assumed
			so that write-allocate transfers cannot be avoided. Note that data transfers
			and code balance pertain to the L3 cache for the stencil codes.} 
		\label{tab:algo}
		\setlength\tabcolsep{0pt}
\setlength\extrarowheight{-0.2pt} 
\begin{tabular*}{\textwidth}{@{\extracolsep{\fill}}llcccwxyzwxyz}
	\specialrule{2.5pt}{1pt}{1pt}
	\rowcolor[gray]{0.9}
	 & {Naive} & {Pseudo-code for} & Elem.\ transf.  & {Code balance} & \multicolumn{4}{c}{Memory request fraction $f$} & \multicolumn{4}{c}{Saturated bandw.\ $b_s$ [\GBS]}	\\
	\rowcolor[gray]{0.9}
	 \multirow{-2}{*}{\rotatebox{90}{\textbf{Cases}}}& {Kernels} & {loop body} & (R+W+RFO)\mbox{$^\ast$} & { $B_c$ [\si{\bytes\per\F}]}  & {BDW-1} & {BDW-2} & {  CLX  } & {Rome} & {BDW-1} & {BDW-2} & {  CLX  } & {Rome}\\
	
	\specialrule{2.5pt}{1pt}{1pt}
	
							\cellcolor[gray]{0.9}& \cellcolor{Cyan!10!white} vectorSUM  & s += a[i] & 1 (1+0+0) & 8  & 0.241  & 0.178 & 0.125& 0.590 
							& 59&66.9&111.1&34.7 \\ 
							\cline{2-9}
							\cellcolor[gray]{0.9}& \cellcolor{Cyan!10!white} DDOT1  & s += a[i]*a[i] & 1 (1+0+0) & 4  &  0.242  & 0.179 & 0.126 & 0.571  
							& 59&66.7&110.5&34.7 \\ \cline{2-9}
							\cellcolor[gray]{0.9}& \cellcolor{Cyan!10!white} DDOT2  & s += a[i]*b[i] & 2 (2+0+0) & 8  & 0.252  & 0.181 & 0.142 & 
							0.665  & 56.5& 65.8&108.7&33.6 \\ \cline{2-9}
							\multirow{-4}{*}{\cellcolor[gray]{0.9}\rotatebox{90}{\textbf{Read-only}}	}& \cellcolor{Cyan!10!white} DDOT3  & s += a[i]*b[i]*c[i] & 3 (3+0+0) & 8  & 0.255  & 0.181  & 0.166 &  0.721 
							& 56.8&65.5&100.9&33.1\\ 
	
	\specialrule{1.65pt}{0.15pt}{0.15pt}
	\multirow{9}{*}{} 
							& \cellcolor{Cyan!10!white} DSCAL & a[i] = s * a[i] & 2 (1+1+0) & 16  & 0.374 & 0.301  & 0.211 & 0.857  
							& 49.6&54.1&101.1&34.9 \\ \cline{2-9}
							& \cellcolor{Cyan!10!white} DAXPY & a[i] = a[i] + s * b[i] & 3 (2+1+0) &12  & 0.315 & 0.239 & 0.204 & 0.960 
							& 53.2&60.8&102.5&32.6 \\ \cline{2-9}
							& \cellcolor{Cyan!10!white} ADD & a[i] = b[i] + c[i] & 4 (2+1+1) & 32  & 0.309 & 0.228 & 0.199& 0.831 
							 & 53.1&62.2&102&32.2 \\  \cline{2-9}
							& \cellcolor{Cyan!10!white} STREAM  & a[i] = b[i] + s * c[i] & 4 (2+1+1) & 16  & 0.309  & 0.228 & 0.199 & 0.838  
							& 53.2&62.2&102.4&32.2 \\ \cline{2-9}
							\cellcolor[gray]{0.9}& \cellcolor{Cyan!10!white} WAXPBY  & a[i] = r * b[i] + s * c[i] & 4 (2+1+1) & 10.67 &  0.309 & 0.228 & 0.199 &0.842 
							& 53.2&62.2&102.4&32.2\\ \cline{2-9}
							\cellcolor[gray]{0.9}& \cellcolor{Cyan!10!white} DCOPY  & a[i] = b[i] & 3 (1+1+1) & 24 [B/row] & 0.320   & 0.242 & 0.190& 0.803  
							& 53.5&60.9&104.2&32.5\\ \cline{2-9}
							\cellcolor[gray]{0.9}& \cellcolor{Cyan!10!white} Schoenauer  & a[i] = b[i] + c[i] * d[i] & 5 (3+1+1) & 20  & 0.299  & 0.223 & 0.185& 0.859 
							& 53.1& 60.5&101.7&31.7\\ \cline{2-9}
							\cellcolor[gray]{0.9}& \cellcolor{Cyan!10!white} Jacobi-v1  & \mbox{$\mathsection \mathsection$} &  &   &  &  &&  & &&&\\
							\cellcolor[gray]{0.9}& \cellcolor{Cyan!10!white} $LC_\mathrm{L2}$ & \mbox{$\dagger$} & 3  (1+1+1) & 6 &  0.252 & 0.195 & 0.157& 0.749  
							& 53.6&60.9&104.1&32.8\\ 
							&\cellcolor{Cyan!10!white} $LC_\mathrm{L3}$ & \mbox{$\ddagger$}  & 5 (3+1+1) & 10 & 0.141  & 0.104 & 0.100 & 0.542
							 & 53.2&60.5&103.2&32.6 \\ 
							\cline{2-9}
							& \cellcolor{Cyan!10!white} Jacobi-v2  & \mbox{$\mathparagraph$} &  &  &  &  &&  & &&&\\ 
							& \cellcolor{Cyan!10!white} $LC_\mathrm{L2}$ & \mbox{$\dagger$}   & 4 (2+1+1) & 2.46 &  0.247 & 0.188 & 0.167&  0.804 
							& 53.5&62.3&102.9&33.2\\
							\multirow{-13}{*}{\rotatebox{90}{\textbf{Read-write}}}
							& \cellcolor{Cyan!10!white} $LC_\mathrm{L3}$ & \mbox{$\ddagger$}   & 6 (4+1+1) & 3.69 &  0.142 & 0.105 & 0.088& 0.458
							 & 52.9&60.8& 103.2&32.1\\ 
	\specialrule{2.5pt}{1pt}{1pt}
\end{tabular*}

		\begin{tablenotes}
			\small			
			\item \mbox{$\ast$} 
			R: no.\ of read streams, 
			W: no.\ of write streams, 
			RFO: read-for-ownership (a.k.a. write allocate).

			\item \mbox{$\mathsection \mathsection$} Simple 2d 5-point stencil update:
			{b[j][i] = (a[j][i-1] + a[j][i+1] + a[j-1][i] + a[j+1][i]) * s}
			
			\item \mbox{$\dagger$} CL transfers and code balance in L3 for layer condition fulfilled at L2 (three data streams); grid size $20000 \times 4000$ (outer $\times$ inner)
			
			\item \mbox{$\ddagger$} CL transfers and code balance in L3 for violated layer condition at L2 (five data streams); grid size $5000 \times 25000$ (outer $\times$ inner)
			
			\item \mbox{$\mathparagraph$}
			More complicated 2d 5-point stencil update:\\
			\hspace*{5mm}\parbox{0.9\textwidth}{r1 = ( ax * (A[j][i-1] + A[j][i+1]) + ay * (A[j-1][i] + A[j+1][i]) + b1 * A[j][i] - F[j][i]) / b1 \\ 
				B[j][i] = A[j][i] - relax * r1 \\ 
				residual += r1 * r1}
		\end{tablenotes}
	\end{threeparttable}
\end{table*}
\section{Hardware and software setup, experimental methodology}\label{sec:exp}
To ensure the broad applicability of our results,
we conducted all experiments on one ccNUMA domain of four different
processors from Intel and AMD (see Table~\ref{tab:system} for
details on the hardware and software setup).
Cluster-on-Die (CoD) and Sub-NUMA Clustering (SNC) were
turned off on Intel CPUs (except for the experiments with HPCG
on BDW in Sect.~\ref{sec:motivation}), NPS4 mode (i.e., four ccNUMA domains per
socket) was active on AMD Rome, and SMT threads were ignored, i.e.,
each application thread was assigned its own physical core.
We always employed the widest
SIMD instruction set supported by the architecture (AVX2
except on CLX where it was AVX-512) and made sure
that the compiler used standard (non-streaming) stores.
The clock speed (core and uncore) was fixed to the base value via the \verb.likwid-setFrequencies.
tool from the LIKWID~\cite{LIKWID,likwidweb} tool suite.

The transparent huge pages (THP) setting was put to ``always,'' and
the NUMA balancing feature was turned off in order to reduce the
performance impact from these settings~\cite{Alappat:ISC20}. All
prefetching mechanisms in the hardware were enabled.

In order to run two groups of threads of different size and control
thread-core affinity we used the \verb.likwid-mpirun. tool.
Working sets were chosen large enough to not
fit into any cache (i.e., at least 10$\times$ the last-level cache
size), and data sharing across overlapping kernels was ruled out
to eliminate the possibility of cache reuse. 

Runtime traces were visualized using
\ac{ITAC}\footnote{\url{https://software.intel.com/en-us/trace-analyzer}}. All
timings were taken using the \CPP{} high-resolution
\verb.Chrono. clock. Individual kernel executions were repeated at least
15 times to even out variations in runtime.
Memory bandwidths were measured by taking the ratio of data
volume and wall-clock time.

The HPCG benchmark results presented in the introduction were obtained
with version 3.1, a problem size of $160^3$ (2.7\,\GB\ per rank), and
a configured runtime of 1800 seconds. In addition to the standard
compiler flags, we used \texttt{-DHPCG\_NO\_OPENMP} for pure MPI,
\texttt{-DHPCG\_CONTIGUOUS\_ARRAYS} for contiguous data layout, and
\texttt{-trace -tcollect -tcollect-filter func.txt -qopt-report} to
enable trace data collection and analysis of selected user functions
(configured in \texttt{func.txt}).

Further details can be obtained from the example code we put up
for download at 
\url{http://tiny.cc/ISPASS-OBS}.

\section{Code and performance metrics}\label{sec:metrics}

The \emph{computational intensity} of a loop is the ratio of
arithmetic operations (``true work,'' e.g., \flops) to the number of
bytes that must be transferred over a data bottleneck in order to
do the work. It is the main parameter that goes into the
\Rlm~\cite{roofline:2009}, which assumes only two hardware
bottlenecks: memory bandwidth and peak computational performance.
Its inverse, the \emph{code balance}, appears at first sight
to be a reasonable metric to quantify the
``hunger'' of a loop for data. However, from the point of view of
refined performance models such as the ECM
model~\cite{sthw15,Hofmann:2020:SF}, arithmetic instructions are an
overlapping component of code execution.  This means that the
execution time of an executing loop does not depend on the actual
number of \flops\ but on the time spent moving data through the memory
hierarchy unless the arithmetic work becomes dominant.  The
\emph{performance} of the loop in \FS\ does depend on the number of
\flops, however, but this is not the observable metric we are
concerned with here.

We cannot give a full introduction to the ECM model here; instead,
we describe the components that are relevant for our analysis
on systems where the memory bandwidth is the sole data transfer
bottleneck on a ccNUMA domain. The following contributions
go into a single-core ECM runtime prediction:
\begin{itemize}
\item $T_{OL}$: In-core execution time of loop instructions that are not
  contributing to $T_{L1Reg}$
\item $T_{L1Reg}$: Optimal execution time of all load and/or store instructions,
  assuming full pipeline throughput; i.e., minimum number of cycles
  required to retire these instructions
\item $T_{Mem}$ Time for all required data transfers over the memory
  interface assuming that the full saturated memory bandwidth can be utilized
\item $\{T_{i}\}$ Times for data transfers over all relevant data
  paths in the cache hierarchy, i.e., L2-L3 and L1-L2 in a three-level
  cache hierarchy
\end{itemize}
The ECM \emph{machine model} makes assumptions about how these contributions
have to be put together to arrive at a single-core runtime
contribution. For example, on all Intel server CPUs to date,
only loads count towards $T_{L1Reg}$, and the runtime is
\bq\label{eq:ecm}
T_{ECM}=\max\left(T_{OL},T_{Mem}+\sum_{i=1}^nT_i+T_{L1Reg}\right)\cma
\eq
i.e., data transfers are nonoverlapping while all instructions
executed (except loads) in the core overlap with everything.
The ECM \emph{application model} provides information about the actual
code (instructions and their dependencies) and data transfer
volumes. It is usually obtained by static code analysis and
assumptions about data transfers. In case of pure streaming
loop kernels without temporal reuse, it is straightforward
to calculate the data volumes. For stencils, a layer condition (LC)
analysis~\cite{sthw15} provides this information. 
\renewcommand{\thefigure}{5}
\begin{figure*}
	\centering
	\begin{minipage}[c]{0.34\textwidth}
		\caption{Basic model of memory bandwidth sharing. In the example, two kernels
			generate requests on six and four cores, respectively,
			and kernel \RN{2} (on MPI rank 1) has
			a much higher $f^\mrn 2\gg f^\mrn 1$.
			It can queue more requests per core and thus get more
			share of bandwidth per core.}
		\label{fig:model}
	\end{minipage}\hfill
	\begin{minipage}[c]{0.64\textwidth}
		\resizebox{\columnwidth}{!}{%
			\input{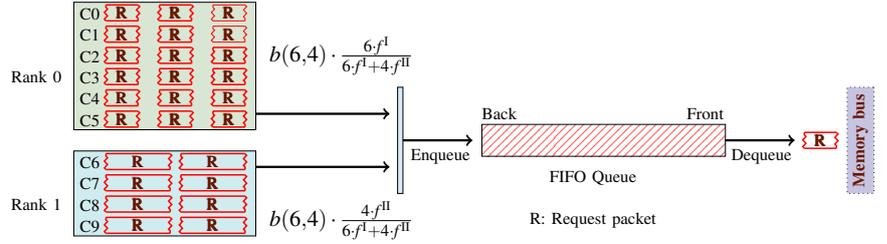}%
		}
	\end{minipage}
\end{figure*}
The ECM model provides us with a resource-centric point of view
on memory bandwidth utilization. The fraction of time
the serial code occupies the memory interface is
\bq
f = \frac{T_{Mem}}{T_{ECM}}\eos
\eq
On a largely nonoverlapping memory hierarchy like in Intel server
CPUs, the \emph{memory request fraction} $f$
is significantly smaller than one even for streaming kernels
with no temporal locality; on AMD Rome, which
has strongly overlapping characteristics, it is often close to one
in this case.
The value of $f$ quantifies how much of the shared resource
(the memory interface) can be utilized by a single core. In
most memory-bound loops, $f$ does not change if the number
of \flops\ (and hence the code balance) changes
because the data transfers dominate in (\ref{eq:ecm}).
This is why code balance is not a good metric
for quantifying the need of a loop for data. In contrast
to $f$, it does not take machine characteristics like
the overlapping behavior of the cache hierarchy into account.

The value of $f$ can be determined by using performance counters
to measure the memory bandwidth drawn when executing a loop
kernel in single-threaded mode. It can also be predicted
by the analytic ECM model if the required knowledge about
the hardware and code (see above) is available. 
We use the first option here and determine $f$ per kernel as
\bq
f = \frac{b_\mathrm{meas}}{b_\mathrm s}\cma
\eq
where $b_\mathrm{meas}$ is the measured single-threaded memory bandwidth
and $b_\mathrm s$ is the saturated (full-domain) memory bandwidth
of the kernel.

The ECM model also predicts the scaling behavior of loops across
cores on a contention domain. We use a simplified version of the
recursive model presented in~\cite{Hofmann:2018}. It assumes that,
at $n$ cores, a latency penalty of $p_0\times u(n-1)\times (n-1)$
must be added, with $u(i)$ being the utilization of the memory interface
at $i$ cores, $u(1)=f$, and $p_0=T_{Mem}/2$. This is not as accurate
as the full model, in which $p_0$ is a fit parameter, but it will
suffice for our purposes.

The maximum memory bandwidth $b_s$ is not the same for all
loop kernels. As a general rule on all x86 CPUs, read-only
kernels achieve a somewhat (5\%--15\%)  higher saturated bandwidth than
kernels with write streams. While this effect is of
minor importance in single-core modeling since the memory data
transfer only accounts for a fraction ($f$) of the execution
time at least on Intel CPUs, an accurate model of bandwidth sharing
requires to take the differences in $b_s$ into account.

Table~\ref{tab:algo} lists the loop kernels used for our experiments together with their
basic properties such as number of elements transferred over the memory interface per iteration
(except for the stencils, where we show transfers to/from the L3 cache), their (memory or L3) code
balance, their $f$ values, and their saturated bandwidths.
We include two 2d stencil algorithms with different characteristics.
For each we select two different grid sizes that exhibit layer conditions (LC) fulfilled
or broken at the L2 cache; if the LC is fulfilled, the data transfers between L3 and L2
are reduced because reuse across the outer stencil dimension is possible at the L2 cache.
This happens when three consecutive rows of the source grid fit into L2.
For the stencils we thus list the code balance in L3 instead of in memory, because
the in-memory code balance is the same regardless of the LC at L2.
However, the intra-cache data transfers between L2 and L3 make a significant
difference for the memory request fraction $f$. See~\cite{sthw15} for an in-depth
coverage of layer conditions.

\renewcommand{\thefigure}{6}
\begin{figure*}[!bt]
	\input{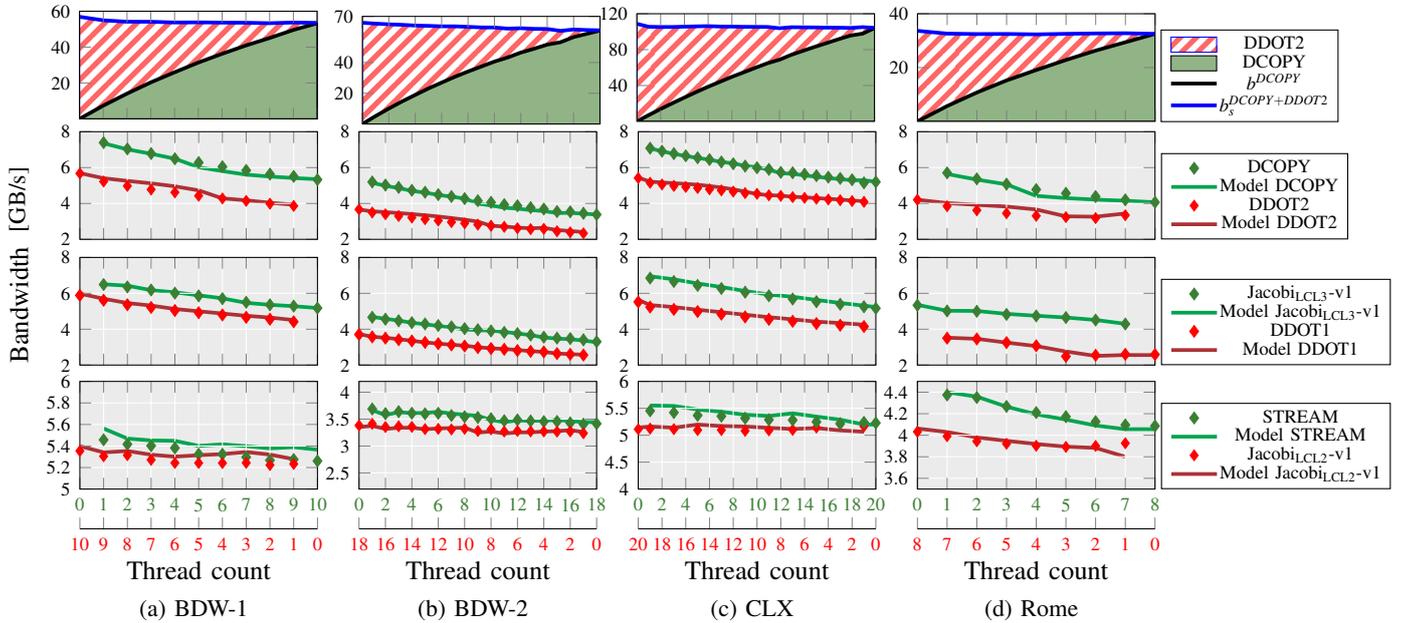}
	\caption{Observed bandwidth per kernel vs.\    
		number of threads per kernel for the fully populated ccNUMA
		domain (thread parameter space: orange dots in Fig.~\ref{fig:threads})
                across all four architectures (columns
		(a)--(d)). Top row: Stacked graph of bandwidth
		share per kernel (DCOPY+DDOT2); the sum is the overall
		memory bandwidth (blue line). Second row: comparison of model (lines)
		with measurement (symbols) for the per-core memory
		bandwidth using the same pairing as above.
		Third and fourth row: same comparison for JacobiL3-v1+DDOT1
		and STREAM+JacobiL2-v1, respectively.
		\label{fig:hybridBench_hetro}}
\end{figure*}
\section{Analytical model}\label{sec:model}
The central quantities in our bandwidth sharing model are the kernel's
memory request fraction $f$ and the saturated bandwidth $b_s$. 
As an example we look at a ten-core contention domain with six cores
running kernel I ($f^\mathrm I$) and four cores running kernel II
($f^\mathrm{II}$) as depicted in Fig.~\ref{fig:model}. Each core
issues requests at a certain fraction of the maximum rate; this
fraction is $f^\mathrm{\RN{1}}$ or $f^\mathrm{\RN{2}}$, respectively.
In the model, this is also the rate at which requests are
queued to be serviced by the memory interface. A kernel with
higher $f$ will be able to queue more requests. Since we are
dealing with a fully populated contention domain, the requests
from the ten cores compete for bandwidth, which is limited
by a value that depends (though weakly) on the code characteristics.
This variation is phenomenological input to the model.
In Table~\ref{tab:algo} we listed the saturation bandwidths $b_s$ for
all kernels at homogeneous execution (no mixing). When two different
kernels overlap, this value changes, and we assume that the
overlapped saturated bandwidth is a weighted mean of
homogeneous bandwidths:
\bq
b(n_t^\mrn{1},n_t^\mrn{2}) = \frac{n_t^\mrn{1}\times b_s^\mrn{1}+n_t^\mrn{2}\times b_s^\mrn{2}}{n_t^\mrn{1}+n_t^\mrn{2}}
\eq
Here, $ b_s^\mrn{1}$ and  $b_s^\mrn{2}$ are the saturated bandwidths
for the two kernels in homogeneous execution, and
$n_t^\mrn{1}$ and $n_t^\mrn{2}$ are the respective numbers of threads,
where $n_t=n_t^\mrn{1}+n_t^\mrn{2}$.

As shown in Fig.~\ref{fig:model}, where we chose $n_t^\mrn{1}=6$ and $n_t^\mrn{2}=4$,
we now assume that group \RN{1} of cores gets a share of all requests,
and thus a share of the bandwidth, proportional to
\bq\label{eq:bsmodel}
\alpha^\mrn 1 =
\frac{n_t^\mrn{1}\times f^\mrn 1}{n_t^\mrn{1}\times f^\mrn 1+n_t^\mrn{2}\times f^\mrn 2}
\quad\mbox{and}\quad\alpha^\mrn 2 =1-\alpha^\mrn 1\eos
\eq
In the homogeneous case $f^\mrn 1=f^\mrn 2$, the share of bandwidth is
solely determined by the number of threads in each group,
but different request fractions modify this simple linear
behavior. The bandwidth obtained by group \RN{1} is thus
$\alpha^\mrn 1\times b(n_t^\mrn{1},n_t^\mrn{2})$.

This model, although we have derived it starting from the
assumption that the ccNUMA domain is fully populated, can also
be applied to the nonsaturated case. In the following section
we will validate the model using measurements in the
parameter space depicted in Fig.~\ref{fig:threads}.

\renewcommand{\thefigure}{7}
\begin{figure*}[bt]
  \input{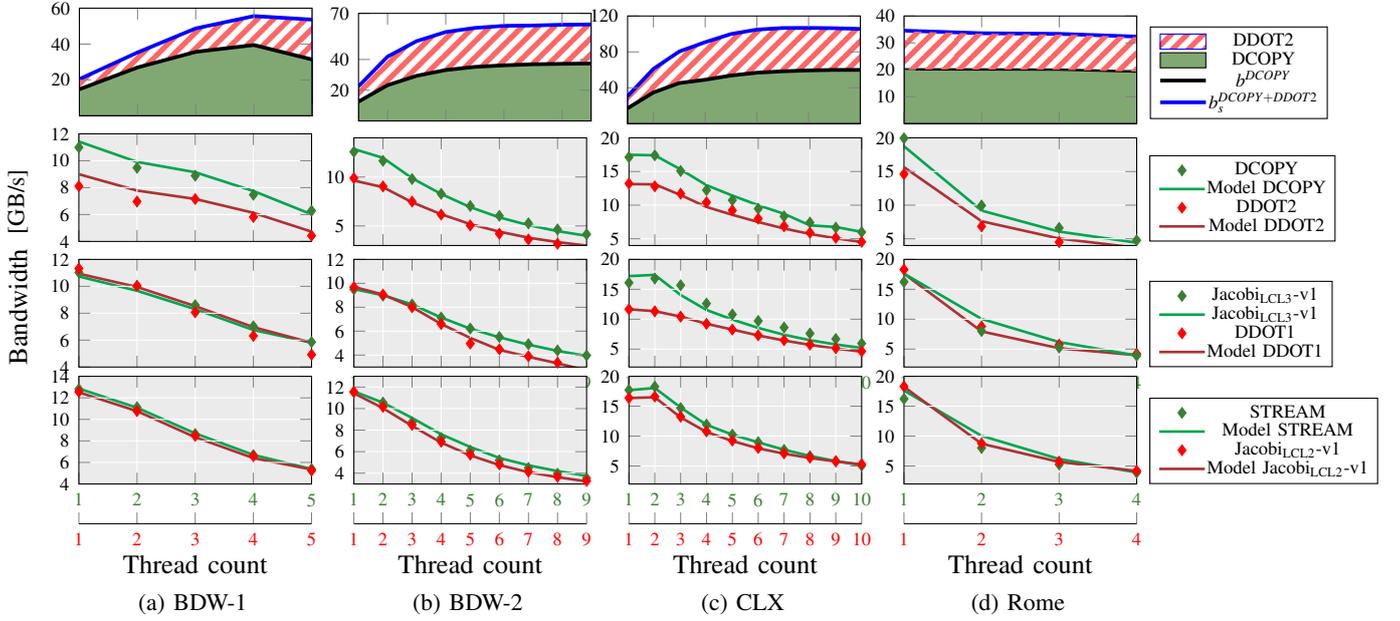}
  \caption{Observed memory bandwidth per kernel vs.\ number of threads
    in symmetrical configuration, scaling across the ccNUMA domain (thread parameter space: blue dots in
    Fig.~\ref{fig:threads}) for all architectures (columns (a)--(d)).
    Organization, kernels, and data plotted as in Fig.~\ref{fig:hybridBench_hetro}.
  \label{fig:hybridBench_homo}}
\end{figure*}

\renewcommand{\thefigure}{8}
\begin{figure*}[!b]
  \centering
\begin{minipage}{\textwidth}
	\vspace{-0.8em}
	\begin{subfigure}[t]{0.254\textwidth} 
		\begin{tikzpicture}
		\pgfplotstableread{figures/fig9/Meggie_boxplot.txt}\MeggieBoxplot;
		\pgfplotstabletranspose\datatransposed{\MeggieBoxplot} 
	
		\begin{axis}[
		boxplot/draw direction = y,
		axis background/.style={fill=white!95!black},
		width=1.18\textwidth,height=0.15 \textheight,
		x label style={at={(0.5,-0.21)}},
		y label style={at={(0.1,0.4)}},
		x tick label style={font=\scriptsize,OliveGreen},
		y tick label style={font=\scriptsize},
		ylabel = {Percentage error},
		y unit= \%,
		xlabel = {Thread count},
		xmin=0,xmax=6,
		ymin=0,ymax=10,
		xtick={1,2,3,4,5},
		xticklabels={1,2,3,4,5},
		axis on top,
		legend style = {
			anchor=south, 
			font=\scriptsize,
			at={(2,1.1)}
		},
		legend columns = 5,
		scaled x ticks=false
		]

\foreach \n in {1,...,5} {
	\addplot+[boxplot, /pgfplots/boxplot/hide outliers,fill, draw=black] table[y index=\n] {\datatransposed};
}
	
		\end{axis}
		\put(0,-15){\begin{axis}[
			axis y line*=none,
			axis x line*=bottom,
			hide y axis,
			tick align = inside,
			x tick label style={font=\scriptsize,red},
			ymin=6, ymax=14,
			xmin=0, xmax=6,
			xtick={1,2,3,4,5},
			width=1.18\textwidth,height=0.15 \textheight,
			]
			\end{axis}}
		\node [font=\small] at (1.5,-1.6){(a) BDW-1}; 
		\end{tikzpicture}
	\end{subfigure}
	\hspace{-1.6em}
	\begin{subfigure}[t]{0.254\textwidth} 
		\begin{tikzpicture}
		
		\pgfplotstableread{figures/fig9/BDW_boxplot.txt}\BDWBoxplot;
		\pgfplotstabletranspose\BDWtranspose{\BDWBoxplot}
		
		\begin{axis}[
		boxplot/draw direction = y,
		axis background/.style={fill=white!95!black},
		width=1.18\textwidth,height=0.15 \textheight,
		x label style={at={(0.5,-0.21)}},
		x tick label style={font=\scriptsize,OliveGreen},
		y tick label style={font=\scriptsize},
		xlabel = {Thread count},
		xmin=0,xmax=10,
		ymin=0,ymax=10,
		xtick={1,2,3,4,5,6,7,8,9},
		axis on top,
		legend style = {
			nodes={inner sep=0.04em}, 
			anchor=south, 
			font=\scriptsize,
			at={(4.9,0)}
		},
		legend columns = 1,
		scaled x ticks=false
		]

\foreach \n in {1,...,9} {
	\addplot+[boxplot, /pgfplots/boxplot/hide outliers, fill, draw=black!50!white] table[y index=\n] {\BDWtranspose};
}

		\end{axis}
		\put(0,-15){\begin{axis}[
			axis y line*=none,
			axis x line*=bottom,
			hide y axis,
			tick align = inside,
			x tick label style={font=\scriptsize,red},
			ymin=6, ymax=14,
			xmin=0, xmax=10,
			xtick={1,2,3,4,5,6,7,8,9},
			width=1.18\textwidth,height=0.15 \textheight,
			]
			\end{axis}}
		\node [font=\small] at (1.5,-1.6){(b) BDW-2}; 
		\end{tikzpicture}
	\end{subfigure}
	\hspace{-1.6em}
	\begin{subfigure}[t]{0.254\textwidth} 
		\begin{tikzpicture}
		\pgfplotstableread{figures/fig9/CLX_boxplot.txt}\CLXBoxplot;
		\pgfplotstabletranspose\CLXtranspose{\CLXBoxplot}
		\begin{axis}[
		boxplot/draw direction = y,
		axis background/.style={fill=white!95!black},
		width=1.18\textwidth,height=0.15 \textheight,
		x label style={at={(0.5,-0.21)}},
		x tick label style={font=\scriptsize,OliveGreen},
		y tick label style={font=\scriptsize},
		xlabel = {Thread count},
		xmin=0,xmax=11,
		ymin=0,ymax=10,
		xtick={1,2,3,4,5,6,7,8,9,10},
		axis on top,
		legend style = {
			nodes={inner sep=0.04em}, 
			anchor=south, 
			font=\scriptsize,
			at={(4.9,0)}
		},
		legend columns = 1,
		scaled x ticks=false
		] 

		\foreach \n in {1,...,10} {
			\addplot+[boxplot, /pgfplots/boxplot/hide outliers,fill, draw=black!50!white] table[y index=\n] {\CLXtranspose};
		}
		\end{axis}
		\put(0,-15){\begin{axis}[
			axis y line*=none,
			axis x line*=bottom,
			hide y axis,
			tick align = inside,
			x tick label style={font=\scriptsize,red},
			ymin=6, ymax=14,
			xmin=0, xmax=11,
			xtick={1,2,3,4,5,6,7,8,9,10},
			width=1.18\textwidth,height=0.15 \textheight,
			]
			\end{axis}}
		\node [font=\small] at (1.5,-1.6){(c) CLX}; 
		\end{tikzpicture}
	\end{subfigure}
	\hspace{-1.6em}
	\begin{subfigure}[t]{0.254\textwidth} 
		\begin{tikzpicture}
		\pgfplotstableread{figures/fig9/Rome_boxplot.txt}\RomeBoxplot;
		\pgfplotstabletranspose\Rometranspose{\RomeBoxplot}
		
		\begin{axis}[
		boxplot/draw direction = y,
		axis background/.style={fill=white!95!black},
		width=1.18\textwidth,height=0.15 \textheight,
		x label style={at={(0.5,-0.21)}},
		x tick label style={font=\scriptsize,OliveGreen},
		y tick label style={font=\scriptsize},
		xlabel = {Thread count},
		xmin=0,xmax=5,
		ymin=0,ymax=10,
		xtick={1,2,3,4},
		axis on top,
		legend style = {
			nodes={inner sep=0.04em}, 
			anchor=south, 
			font=\scriptsize,
			at={(4.9,0)}
		},
		legend columns = 1,
		scaled x ticks=false
		] 
		\foreach \n in {1,...,4} {
			\addplot+[boxplot, /pgfplots/boxplot/hide outliers,fill, draw=black!50!white] table[y index=\n] {\Rometranspose};
		}		
		
		\end{axis}
			\put(0,-15){\begin{axis}[
			axis y line*=none,
			axis x line*=bottom,
			hide y axis,
			tick align = inside,
			x tick label style={font=\scriptsize,red},
			ymin=6, ymax=14,
			xmin=0, xmax=5,
			xtick={1,2,3,4},
			width=1.18\textwidth,height=0.15 \textheight,
			]
			\end{axis}}
		\node [font=\small] at (1.5,-1.6){(d) Rome }; 
		\end{tikzpicture}
	\end{subfigure}
\end{minipage}
  \caption{Overview of modeling error across architectures (a)--(d)
    for symmetrical thread scaling (blue dots in Fig.~\ref{fig:threads})
    and 30 pairings per thread count and architecture. 
    The error is calculated as
    $\left\lvert(b_\mathrm{Observed}-b_\mathrm{Model})/b_\mathrm{Model}\right\rvert$.
    Whiskers denote min/max error, boxes encompass second and third quartiles,
    and the median is marked. 
  \label{fig:error}}
\end{figure*}
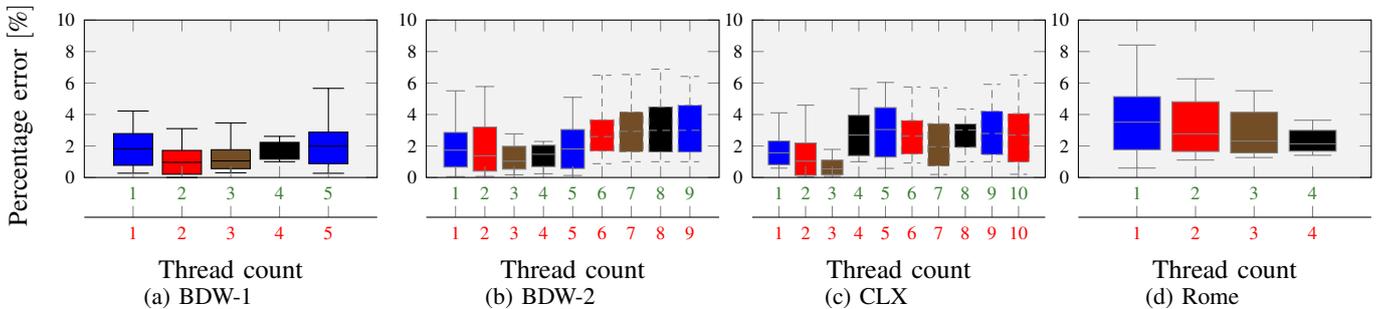

\section{Model validation and discussion}\label{sec:validation}

In this section we first present measurements of selected scenarios
(i.e., kernel pairings) on the four architectures and compare with the
model (\ref{eq:bsmodel}), distinguishing between the fully populated
domain case and the case where the number of cores goes up
till saturation. We then extend the view to more pairings
and finally present an overview of the modeling error for
all cases.

In Fig.~\ref{fig:hybridBench_hetro} we show three different kernel pairing
scenarios per architecture using fully populated ccNUMA domains (i.e., 10, 18, 20, and 8 cores
on the four architectures (a)--(d)), covering the orange dots in Fig.~\ref{fig:threads}. The top
panel in each column shows a stacked graph of the bandwidth
share of kernel \RN 1 (DCOPY, green) versus kernel \RN 2 (DDOT2, hatched red)
as the number of threads on kernel \RN 1 is increased and the number
of threads on kernel \RN 2 is reduced. 
The top line in the graph is the measured overall memory bandwidth.
Since DCOPY has a higher $f$ than DDOT2 (see Table~\ref{tab:algo}),
we expect from the model (\ref{eq:bsmodel}) that DCOPY will
get a higher share of the bandwidth as the number of DCOPY
threads goes up. This is observed as the upward ``bend'' of the
separator between the two regions of the graph. The overall memory
bandwidth goes down because the saturation bandwidth of DCOPY is
smaller than that of DDOT2 (which is a read-only kernel that gets
a little more bandwidth out of the memory interface).
This behavior is quite universal across all four architectures 
for this kernel pairing.

The second panel in each column is a direct comparison between
the modeled (lines) and observed (symbols) memory bandwidth per core
for the DCOPY+DDOT2 case. Since we chose a zoomed-in $y$ axis,
the general downward trend stemming from the decline in saturated
bandwidth is now more pronounced. Although not directly visible
from the mathematical description, this change is just as important
for the observed bandwidth as the difference in $f$. Obviously,
the model describes the per-core bandwidth quite accurately.

The third and fourth panels in each column show per-core
bandwidth for two other pairings: JacobiL3-v1+DDOT1 and
STREAM+JacobiL2-v1. Especially in the latter case the
saturated bandwidths of the two kernels are very similar,
so the downward trend is weaker and also the bandwidth
difference per core is smaller. Nevertheless the model
also provides an accurate prediction here.

Figure~\ref{fig:hybridBench_homo} shows the same cases as above but
for symmetrical thread scaling, running the same number of threads
per kernel (blue dots in Fig.~\ref{fig:threads}). There is good agreement
between model and measurement along the bandwidth saturation
curve as well, but some peculiarities are worth noting. The
notable breakdown in bandwidth for DCOPY near saturation
on BDW-1 is reminiscent of the overall scaling behavior of this
kernel on this architecture, which shows a maximum bandwidth
already before the domain is full. While the BDW CPUs and especially Rome
show a rather strong decline in per-thread bandwidth already at
two threads per kernel, CLX scales well from two to four,
which reflects the fact that its single-core bandwidth
is low compared to its saturated bandwidth. Rome is
special in terms of scaling since all kernels can almost
saturate the memory bandwidth already with one thread due
to its overlapping cache hierarchy. A more complete coverage
of scaling results for more pairings can be found at 
\url{http://tiny.cc/ISPASS-OBS}.

In Fig.~\ref{fig:error} we give a concise overview of the relative
modeling error in per-core bandwidth across 30 kernel pairings
along the bandwidth scaling curve, i.e., for symmetrical pairing.
In 75\% of all cases the error is below 5\%, and the maximum
overall error is 8\%. In view of the significant differences
in architectural details across the four CPUs (inclusive vs.\
victim LLC, shared vs.\ segmented LLC, overlapping vs.\
serializing caches) we consider this an exceptional result.

\renewcommand{\thefigure}{9}
\begin{figure*}[bt]
  \input{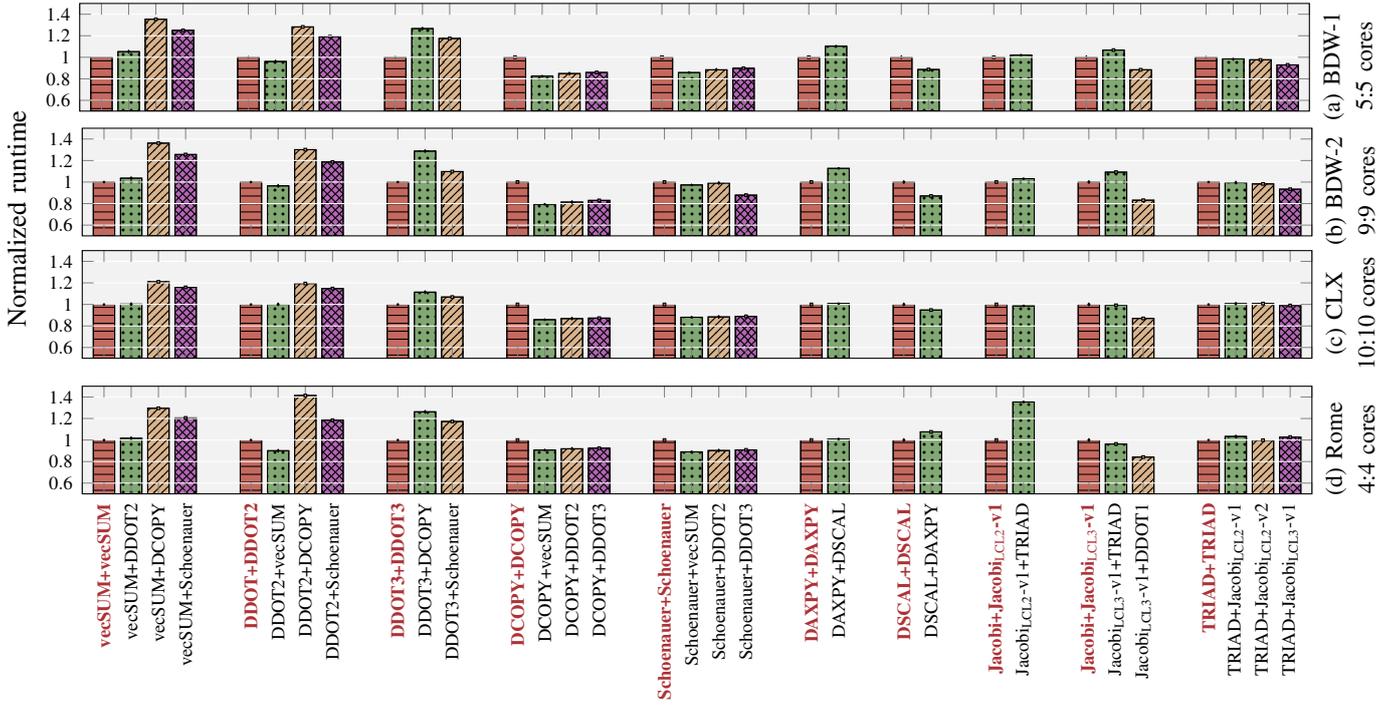}
  \caption{Bandwidth gain or loss for symmetrical kernel pairings
    (each kernel gets half the contention domain) across architectures
    (a)--(d).  The height of each bar is the relative gain or loss in
    bandwidth of the first kernel in the pair when paired
    with the second, normalized to the self-paired case (first
    bar in each group).
  \label{fig:pure_hybridBench}}
\end{figure*}
Figure~\ref{fig:pure_hybridBench} shows an overview of 32 kernel pairings
(including self pairings) of vecSUM, DDOT2, DDOT3, DCOPY, Schoenauer,
DAXPY, DSCAL, JacobiL2-v1, JacobiL3-v1, and TRIAD with others using
an equal share of threads on a full contention domain. Each row
of bars (a)--(d) refers to one architecture. Each bar represents
the relative bandwidth share of the first kernel in the pair (e.g.,
vecSUM in vecSUM+DDOT2) as opposed to the homogeneous situation
(e.g., vecSUM paired with itself). All bars in a group are
thus normalized to the first bar.

The first observation is that the patterns observed in each group of
bars are quite consistent across architectures: Whether bandwidth is
gained or lost with respect to the homogeneous situation depends on
the ratio of $f$ values of the kernels being greater or smaller than
one. Across the Intel CPUs, this criterion is independent of the
architecture. The CLX CPU is still special in the sense that the
differences in bandwidth are smaller overall.  One might be tempted to
attribute this to the fact that it needs more cores to saturate the
memory bandwidth than the BDW variants, i.e., it is ``more scalable.''
In other words, the memory transfers need less time relative to the
other contributions in the single-core ECM model, which cause
generally smaller request fractions $f$ as can be seen in
Table~\ref{tab:algo}.  However, a global reduction factor in $f$
cancels out in the model (\ref{eq:bsmodel}), hence this cannot be the
reason for the weaker bandwidth variations. The real reason for CLX
showing smaller variations is two-fold: first, it shows less spread in
saturated bandwidth than the other Intel CPUs across all kernels in
Table~\ref{tab:algo} (10\% as opposed to 20\% for BDW-1); second, it
also shows less spread in $f$ values (2.4 vs.\ 2.7 for BDW-1).  These
two factors together lead to the reduced sensitivity of shared
bandwidth to kernel variations on CLX.
The AMD Rome CPU shows different patterns from the Intel chips for
some combinations, especially for DAXPY+DSCAL. This due to
$f^\mathrm{DAXPY}>f^\mathrm{DSCAL}$ on Rome as opposed to
the Intel CPUs where this relation is reversed. 

We started this discussion with a close look at plain and modified
MPI-parallel HPCG variants (see Sect.~\ref{sec:motivation}), where we
observed how back-to-back compute kernels overlapping on a contention
domain due to desynchronization could either lead to them slowing down
or speeding up depending on the particular pairing.
If a kernel is sandwiched between a high-$f$ kernel coming before it
and a low-$f$ kernel coming after it, early starters get slowed
down and late starters get sped up. This could be observed
in Fig.~\ref{fig:Modified_HPCG}(b), where the follow-up kernel
to DDOT2 was DAXPY, with $f^\mathrm{DAXPY}=0.315$ and $f^\mathrm{DDOT2}=0.252$.
This large difference directly leads to the positive
observed skewness and means that the desynchronization is amplified
in such a situation. Overlapping with idleness, as shown in
Fig.~\ref{fig:Modified_HPCG}(a) on the trailing edge of the
DDOT2 execution, causes resynchronization. Our model is thus
not only good for a quantitative description of performance
differences on shared bottlenecks; it can also predict
qualitatively the dynamics of desynchronization and resynchronization
in memory-bound bulk-synchronous barrier-free programs.

\section{Conclusion and outlook}\label{sec:conclusion}

Starting from observations of desynchronized kernel execution on
the MPI-parallel HPCG benchmark we motivated the need for an
analytical model of bandwidth sharing between groups of threads
executing different loop kernels with different characteristics
on a bandwidth contention domain.
Based on the principles of the Execution-Cache-Memory (ECM)
performance model, we constructed and validated such a model
on four current x86 server processors from Intel and AMD.
We could show that the
major influence factors for bandwidth sharing between two groups
of threads are each kernel's memory request frequency, which is
directly related to its single-threaded memory bandwidth,
and its saturated bandwidth on the domain. Across a variety
of pairings of kernels with and without cache reuse,
the observed error in predicted per-core bandwidth
was never larger than 8\%, and lower than 5\% for 75\% of all
cases. Apart from the quantification of bandwidth
shares, the model is also able to predict how desynchronization
across MPI processes gets amplified or mitigated depending on
which back-to-back kernels are overlapped with each other
in MPI-parallel barrier-free bulk-synchronous programs. 

The set of loop kernels chosen here is a reasonable cross section with
a spectrum of properties. Since the memory request fraction and the
saturated memory bandwidth are the only relevant parameters in our
model, it should be applicable also for more complex kernels (e.g.,
with more concurrent data streams or with dominant in-core execution).
It should also be useful in modeling the performance of task-parallel
code where the synchronized, data-parallel execution of threads may be
more the exception than the rule. A validation of the model on more
processor architectures, e.g., Power- or Arm- based CPUs, is certainly
in order. Finally, our work enables the development of a new kind of
MPI simulation technique that can take node-level bottlenecks into
account much more accurately than previously possible. We leave these
investigations for future work.

\ifblind
\else
\section*{Acknowledgments}
This work was supported by KONWIHR, the Bavarian Competence Network
for Scientific High Performance Computing in Bavaria, under project name ``OMI4papps.''
\fi

\printbibliography
 
\end{document}